\documentclass[11pt, oneside]{article}

\usepackage{geometry}
\usepackage{graphicx}
\usepackage{amssymb}
\usepackage{bm}
\usepackage{here}
\usepackage{geometry}
\usepackage{natbib}
\usepackage{url}
\usepackage{breakurl}
\usepackage[title]{appendix}
\usepackage{booktabs}
\usepackage{multirow}
\usepackage{siunitx}

\usepackage{amsmath}

\usepackage{lscape}
\usepackage{comment}

\usepackage{subcaption}

\title{A growth model for water distribution networks with loops}

\author{Kashin Sugishita and Naoki Masuda}

\usepackage{color}

\begin{document}

\begin{center}
    {\LARGE\bf A growth model for water distribution networks\\ with loops}
\end{center}
\begin{center}
    Kashin Sugishita$^{1,2}$, Noha Abdel-Mottaleb$^{3}$,  Qiong Zhang$^{3}$, and Naoki Masuda$^{1,4,5}$\\
\end{center}
%
%
{\small
$^{1}$Department of Mathematics, State University of New York at Buffalo, Buffalo, NY 14260-2900, USA\\
$^{2}$Department of Transdisciplinary Science and Engineering, Tokyo Institute of Technology, 152-8550 Tokyo, Japan\\
$^{3}$Department of Civil and Environmental Engineering, University of South Florida, Tampa, FL 33620, USA\\
$^{4}$Computational and Data-Enabled Science and Engineering Program, State University of New York at Buffalo, Buffalo, NY 14260-5030, USA\\
$^{5}$Faculty of Science and Engineering, Waseda University, 169-8555 Tokyo, Japan}\\


\begin{abstract}
Water distribution networks (WDNs) expand their service areas over time. These growth dynamics are poorly understood. One facet of WDNs is that they have loops in general, and closing loops may be a functionally important process for enhancing their robustness and efficiency. 
We propose a growth model for WDNs which generates networks with loops and is applicable to networks with multiple water sources. 
We apply the proposed model to four empirical WDNs to show that it produces networks whose structure is similar to that of the empirical WDNs. The comparison between the empirical and modeled WDNs suggests that the empirical WDNs may realize a reasonable balance between cost, efficiency, and robustness in terms of the network structure. 
We also study the design of pipe diameters based on a biological positive feedback mechanism. Specifically, we apply a model inspired by \textit{Physarum polycephalum} to find moderate positive correlations between the empirical and modeled pipe diameters. The difference between the empirical and modeled pipe diameters suggests that we may be able to improve the performance of WDNs by following organizing principles of biological flow networks.

\end{abstract}

\section{Introduction}\label{sec:introduction}
Water distribution networks (WDNs) are a critical infrastructure. A WDN consists of interconnected hydraulic elements such as pipes, junctions, reservoirs, and tanks, and conveys water from sources to consumers \cite{eiger1994optimal}. A WDN naturally forms a graph in which the nodes represent junctions, sources, control elements, and consumption points, and the edges represent pipes connecting pairs of nodes. For assessing the performance of existing WDNs and designing efficient WDNs, we should first understand structural and hydraulic properties of real-world WDNs and mechanisms behind their genesis. 

Like other infrastructure networks such as transportation networks, power grids, and gas pipeline networks, WDNs are embedded in space. Previous studies showed that spatial networks are different from other complex networks that are not associated with space \cite{barthelemy2011spatial,barthelemy2018morphogenesis}. In particular, one can naturally measure the cost for spatial networks by the length of edges, and the cost of edges generally constrains the structure of spatial networks \cite{barthelemy2011spatial}. For example, the degree (i.e., the number of edges that a node has) distribution of spatial networks is narrow and peaked at small values, which contrasts to fat-tailed degree distributions observed for various non-spatial empirical networks \cite{cardillo2006structural, lammer2006scaling}. In spatial networks, the degree correlation is generally weak because the physical distance between nodes rather than the degree of nodes usually governs the existence of edges in spatial networks \cite{barrat2005effects,sen2003small}. Spatial constraints also induce large fluctuations in the betweenness centrality \cite{guimera2004modeling, guimera2005worldwide}.

Though water distribution networks evolve and expand over time, there is a lack of generative network models that capture how real-world WDNs emerge and develop. A barrier to the development of generative models may be that publicly available information on real-world WDNs is limited for security reasons. Existing generative models for WDNs fall into two categories. The first category lacks relevant urban context (e.g., location of reservoirs and demand junctions). In particular, this category of models produces networks that cannot be directly compared to the structure of empirical WDNs. For example, the Modular Design System model can generate WDNs in which all the pipes have the equal length \cite{moderl2011systematic}. The WaterNetGen model can generate WDNs in which nodes are distributed uniformly at random in a rectangular space \cite{muranho2012waternetgen}. The HydroGen model can generate WDNs with clusters in each of which nodes (e.g., junctions, reservoirs, and tanks) are distributed uniformly at random in a disc \cite{de2014hydrogen}. In Ref. \cite{zeng2017modeling}, the authors propose a generative model for WDNs in which nodes are distributed uniformly at random in a disc and an additional water source node is located at the center of the disc. These models do not intend to mimic the structure of empirical WDNs or the spatial configuration of their nodes and pipes.

The second category of models uses urban setting data but does not consider urban development over time. These models produce realistic network structure because they account for the location of reservoirs and demand junctions, and limited options for edges given existing road networks. In Ref. \cite{sitzenfrei2013automatic}, the authors propose a generative model for WDNs based on geographic information system (GIS) data. They divided a studied area into cells and selected a small network for each cell from the predefined set of network motifs. In Ref. \cite{mair2014spanning}, the authors proposed a different generative model for WDNs based on GIS data. In this model, one extracts a tree network from street network data and adds water sources and loops. These models are successful in producing networks similar to empirical WDNs \cite{mair2014spanning,sitzenfrei2013automatic}. However, they are static models and do not explain the evolution of empirical WDNs over time. In addition, these models require numerous inputs such as street network data, population density, housing density, and a digital elevation map in the GIS data. In contrast, the model proposed in Ref. \cite{gastner2006shape} does not require detailed inputs apart from the physical location of each node. It is a growth model of networks that produces tree networks that are similar to some empirical infrastructure networks such as gas pipeline networks, sewage networks, and railway networks. However, an important limitation of this model is to the application to WDNs because, in contrast to sewer water networks, empirical WDNs usually have an innegligible amount of loops. Loops in flow networks have been suggested to realize optimal transport efficiency under fluctuations in flow  \cite{corson2010fluctuations,hu2013adaptation,katifori2010damage} and some tolerance to damages \cite{dziedzic2014water, katifori2010damage}. Overall, there is not a generative model of WDNs that produces networks similar in structure to empirical WDNs without relying on detailed geospatial data of the empirical urban setting.

To address this gap, the research objectives of the present study are two-fold. First, we propose a growth model for WDNs that produces networks quantitatively similar to empirical WDNs and do not require detailed input apart from the physical location of each node. The proposed model is classified as a greedy model, in which one adds edges one by one based on a local optimization principle. Our model generates networks with loops and is applicable to networks with multiple root nodes. Note that many empirical WDNs have multiple root nodes (e.g., \textsc{Modena} \cite{bragalli2008water}, \textsc{Colorado} \cite{lippai2005water},  \textsc{Pescara} \cite{bragalli2008water}, and \textsc{Balerma} \cite{reca2006genetic}). 
Second, we aim to understand the genesis of the distribution of pipe diameters in the empirical WDNs. The pipe diameter generally varies across pipes in a WDN. It is not well-known whether the distribution of pipe diameters in empirical WDNs attains approximate optimality or there is a large room for improving the efficiency by engineering the pipe diameters under the given constraints. Specifically, we apply a model that is inspired by an amoeba-like organism, \textit{Physarum polycephalum}, with which one gradually adjusts the conductance of each edge \cite{tero2010rules}, to our empirical and synthetic WDNs. In Ref. \cite{skulovichadaptation}, the authors applied a similar \textit{Physarum polycephalum} model to relatively small benchmark WDNs and compared the wiring cost between the generated and empirical WDNs. \textit{Physarum polycephalum} has shown various intelligent behavior such as finding the shortest paths \cite{nakagaki2000maze}, adapting to changing environments  \cite{tero2007mathematical}, and building high-quality networks that realize a feasible balance between cost, efficiency, and robustness \cite{tero2010rules}. Previous studies suggested that \textit{Physarum polycephalum} may inform the design of next-generation, adaptive, and robust spatial infrastructure networks with decentralized control systems \cite{bebber2007biological,kunita2013adaptive,sun2017physarum}.

\section{Structural properties of water distribution networks}
\label{sec:structural_properties_WDNs}
\subsection{Data}
We analyze four WDNs constructed from the empirical data sets, i.e., \textsc{ZJ}, \textsc{Colorado}, \textsc{Modena}, and \textsc{Tampa}. These networks are WDNs in Zhijiang in China, Colorado Springs in the US, Modena in Italy, and Tampa in the US, respectively. The data sets include the IDs and the physical location of junctions, reservoirs, tanks, and pipes, and the diameter of pipes. We model each WDN as a graph $G=G(V, E)$ in which $V$ is the set of $N$ nodes (i.e., junctions, reservoirs, and tanks) and $E$ is the set of $M$ edges (i.e., pipes). Each WDN has one or more reservoirs and/or tanks, which act as source of water and is referred to as root node. We denote the set of the root nodes by $V_R$. We assume that all the nodes except for the root nodes are demand nodes and denote the set of the demand nodes by $V_D$.

\subsection{Indices to be measured}
In addition to the number of nodes and edges, we measure six indices, two of which are the average degree, denoted by $\langle k \rangle$, and the maximum degree, denoted by $k_{\rm max}$. The other four indices are meshedness coefficient, cost, route factor, and robustness. 

First, the meshedness coefficient, denoted by $m$, quantifies the amount of loops for planar graphs\cite{buhl2006topological} and is defined by
\begin{equation}
m = \frac{M-N+1}{2N-5}\,,
\label{eq:meshedness}
\end{equation}
where $N$ and $M$ are the number of nodes and edges, respectively. It ranges between zero (for tree graphs) and one (for maximal planar graphs). WDNs are generally near-planar graphs\cite{yazdani2011complex}. In fact, among the aforementioned four WDNs, the \textsc{ZJ}, \textsc{Colorado}, and \textsc{Modena} networks are planar graphs. 

Second, we define the cost $c$ as the total Euclidean length of all edges, which is the same definition as that in Ref. \cite{gastner2006shape}. The reason for using this definition is that our growth model can be regarded as an extension of a model proposed in Ref. \cite{gastner2006shape}, which was shown to produce tree-like networks similar to real-world tree-like networks such as sewage networks and gas pipeline networks in terms of the network structure. This definition of the cost is common in studies on spatial networks \cite{barthelemy2018morphogenesis}.  

Third, we define the route factor as a measurement for the network efficiency. We use the definition in Ref. \cite{gastner2006shape}, which is also commonly applied to spatial networks \cite{barthelemy2018morphogenesis}. For spatial networks in which flows are transported from the root nodes to the demand nodes, a network should be more efficient if the paths from the root nodes to each demand node are shorter. For a network with the single root node, denoted by $0$, we define the route factor $q$ by
\begin{equation}
        q = \frac{1}{|V_D|}\sum_{i=1}^{|V_D|} \frac{l_{i0}}{d_{i0}}\,,
        \label{eq:route_factor}
\end{equation}
where $d_{i0}$ is Euclidean distance between the root node and demand node $i$, and $l_{i0}$ is the shortest Euclidean distance of the path among the paths between the root node and demand node $i$ \cite{gastner2006shape}. The route factor is always greater than or equal to one, and a value close to one implies a high efficiency. The route factor for networks with multiple root nodes is defined as the average of the route factor over the root nodes. 

Fourth, for WDNs, it is necessary that each node is connected to a root node for the node to receive water supply. Therefore, we define the robustness $R$ by 
\begin{equation}
    R = \frac{1}{M}\sum_{e=1}^{M} s(e)\,,
    \label{eq:robustness}
\end{equation}
where $s(e)$ is the fraction of demand nodes that are connected to any root node after one removes $e$ edges. A simulation of sequential edge removal starts from the given WDN. At every step of the edge removal process, we remove an edge selected uniformly at random. We calculate $R$ as the average over 100 simulations. This index is an adjustment of the one proposed in Ref.~\cite{schneider2011mitigation} to the case of failures of edges and the connectivity to root nodes.

\subsection{Results}

We show the values of the six indices for the four empirical WDNs in Table \ref{tab:structural_properties_comparison}. We also visualize these networks in Fig.~\ref{fig:network_visualization}(a). In Fig.~\ref{fig:network_visualization}, the red and blue nodes represent the root nodes and the demand nodes, respectively. The \textsc{ZJ} and \textsc{Tampa} networks have one root node. The \textsc{Colorado} and \textsc{Modena} networks have four root nodes. 
Each WDN has just one connected component.

All the four WDNs have similar average degrees, $\langle k \rangle\approx 2.5$. The maximum degree is 4 for \textsc{ZJ}, \textsc{Colorado}, and \textsc{Tampa}, and is 5 for \textsc{Modena}. These results are reasonable because the spatial constraints of networks strongly restrict the degree of each node in general \cite{barthelemy2011spatial,barthelemy2018morphogenesis}. 
The meshedness coefficient, $m$, ranges from 0.0580 to 0.229, which is consistent with the previous observation that $m$ is generally less than 0.5 for WDNs \cite{hwang2017water, yazdani2011complex}.
The cost $c$ ranges between 6.31 for \textsc{ZJ} and 590 for \textsc{Tampa}, which depends on the size of the service areas. The four WDNs have similar values of the route factor, $q\approx 1.3$. Finally, \textsc{Modena} has the highest robustness, followed by \textsc{ZJ}, \textsc{Colorado}, and \textsc{Tampa}. This last result may be related to the number of root nodes per demand node, which is also the highest for \textsc{Modena}, followed by \textsc{ZJ}, \textsc{Colorado}, and \textsc{Tampa}.

\begin{table}[tbh]
\caption{Structural properties of the empirical WDNs, the synthetic networks generated by our growth model, and the synthetic networks generated by the Waxman model. $N$: number of nodes, $|V_R|$: number of root nodes, $M$: number of edges, $J$: Jaccard index for the sets of edges between the empirical and synthetic networks, $N_{\rm c}$: number of connected components, $\langle k \rangle$: average degree, $k_{\rm max}$: maximum degree, $m$: meshedness coefficient, $c$: cost (i.e., total edge length) [km], $q$: route factor, and $R$: robustness.}
\label{tab:structural_properties_comparison}
\begin{tabular}{cccc|ccc}
\hline
           & \multicolumn{3}{c|}{\textsc{ZJ}}  & \multicolumn{3}{c}{\textsc{Colorado}}\\
     & Empirical       & Our model     & Waxman      & Empirical          & Our model          &Waxman     \\
    \hline
$N$          & 114 & 114& 114& 1{,}786& 1{,}786& 1{,}786    \\
$|V_R|$         & 1& 1& 1 & 4& 4& 4                 \\
$M$          & 164     & 167   & 163   & 1{,}992  & 1{,}980  & 2{,}004 \\
$J$          & N/A      & 0.689 & 0.131 & N/A         & 0.513      & 0.0670  \\
$N_{\rm c}$         & 1       & 1     & 6    & 1          & 1          & 258 \\
$\langle k \rangle$          & 2.88    & 2.93  & 2.86  & 2.23       & 2.22       & 2.24 \\
$k_{\rm max}$      & 4       & 4     & 8     & 4          & 5          & 10     \\
$m$          & 0.229   & 0.242 & 0.224 & 0.0580     & 0.0547     & 0.0614\\
$c$          & 6.31    & 5.96  & 12.2  & 373        & 353        & 1{,}209   \\
$q$          & 1.29    & 1.28  & N/A    & 1.45       & 1.36       & N/A    \\
$R$          & 0.216   & 0.352 & N/A    & 0.163      & 0.160      & N/A     \\
References &    \cite{zheng2011combined,ZJ}     & N/A     &N/A      &   \cite{lippai2005water, Colorado_Springs}         &   N/A        &N/A\\        
\hline
\end{tabular}

\begin{tabular}{cccc|ccc}
\hline
           & \multicolumn{3}{c|}{\textsc{Modena}}  & \multicolumn{3}{c}{\textsc{Tampa}}\\
     & Empirical       & Our model     & Waxman      & Empirical          & Our model          &Waxman     \\
    \hline
$N$          & 272 & 272& 272& 1{,}658& 1{,}658& 1{,}658    \\
$|V_R|$         & 4& 4& 4 & 1& 1& 1                 \\
$M$          & 317    & 313   & 319   & 1{,}978  & 1{,}993  & 1{,}967 \\
$J$          & N/A      & 0.628 & 0.156 & N/A         & 0.538      & 0.0708  \\
$N_{\rm c}$         & 1       & 1     & 52    & 1          & 1          & 628 \\
$\langle k \rangle$          & 2.33    & 2.30  & 2.35  & 2.39       & 2.40       & 2.37 \\
$k_{\rm max}$      & 5       & 4     & 8     & 4          & 5          & 19     \\
$m$          & 0.0853   & 0.0779 & 0.0891 & 0.0970     & 0.101     & 0.0936\\
$c$          & 71.4   & 70.8  & 116  & 590       & 600        & 764  \\
$q$          & 1.36   & 1.42  & N/A    & 1.27       & 1.25       & N/A    \\
$R$          & 0.231   & 0.250 &N/A    & 0.132      & 0.163      &N/A     \\
References &    \cite{bragalli2012optimal, Modena}     &N/A     &N/A      & N/A  &   N/A        &N/A\\        
\hline
\end{tabular}

\end{table}

\section{Growth Model}
\label{sec:growth_model}

\subsection{Model}
In this section, we propose a growth model for WDNs. The proposed model is classified as a greedy model where edges are added one by one based on a local optimization criterion \cite{barthelemy2011spatial,barthelemy2018morphogenesis}. Our model generates networks with loops and is capable of
generating networks with multiple root nodes. 

Our model takes the number of nodes $N$, the number of edges $M$, the two-dimensional coordinate of the nodes, and the ID of the root node(s) as input. It has one non-negative parameter $\gamma$. The model consists in the following three steps:
\begin{enumerate}
    \item Initially, all nodes are isolated.
    \item We add an edge to the network as follows. We set $P_{\rm loop}=(M-N+1)/M$. 
\begin{enumerate}
    \item \emph{Expansion of a connected component including a root node}: With probability $1-P_{\rm loop}$, we add an edge between nodes $i$ and $j$, which we select as follows. We impose that node $i$ is in a connected component including a root node and that node $j$ is in a connected component that is different from the one including the node $i$. Under this condition, we select $i$ and $j$ such that the Euclidean distance between $i$ and $j$, denoted by $d_{ij}$, is the smallest among all possible node pairs.
    \item \emph{Loop closure}: Alternatively, i.e., with probability $P_{\rm loop}$, we add an edge to close a loop. We impose that node $i$ has degree one and that node $j$ is in the same connected component as the one that contains node $i$. Under this condition, we select $i$ and $j$ that minimize
        \begin{equation}
        w_{ij} \equiv d_{ij}-\gamma \ell_{ij}\,,
        \label{eq:w_ij}
        \end{equation}
    where $\ell_{ij}$ is the Euclidean distance of the path that is the shortest among the paths between nodes $i$ and $j$. Then, we add an edge between $i$ and $j$.  If the number of nodes in the largest connected component is less than three, we carry out step (a) with probability 1 because we cannot carry out the loop closure without creating a multiple edge (i.e., more than one edges directly connecting two nodes). If there is no node with degree one, we also carry out step (a) with probability 1.
\end{enumerate} 
    \item Repeat step (ii) until all the nodes are connected as one component.
\end{enumerate}

Note that the reason why we set $P_{\rm loop}=(M-N+1)/M$ is that the expected number of edges should be $M$ when we complete steps (i)--(iii). In other words, in order for the $N$ nodes to be connected for the first time when we add $M$ edges, we need to perform step (ii-a) exactly $N-1$ times. In expectation, this condition is equivalent to $M(1-P_{\rm loop})=N-1$, which leads to $P_{\rm loop}=(M-N+1)/M$.

Closing loops has been suggested to help realizing optimal transport efficiency under fluctuations in flow distributions \cite{corson2010fluctuations,hu2013adaptation,katifori2010damage} and improving robustness \cite{dziedzic2014water, katifori2010damage}. In our model, parameter $\gamma$ controls how loops are closed. On the right-hand side of Eq.~\eqref{eq:w_ij}, the quantity $d_{ij}$ represents the cost of the new edge. The quantity $\gamma \ell_{ij}$ represents the gain of closing a path of Euculidean length $\ell_{ij}$ to make a loop. 
In other words, when we connect two nodes with large $\ell_{ij}$, the Euclidean length of the shortest loop closed by the new edge is large. 
Parameter $\gamma$ controls the trade-off between the Euclidean length of the new edge and the Euclidean length of the shortest path closed by the new edge. With $\gamma=0$, we always add the shortest edge in terms of the Euclidean distance. As we will show later, the generated networks with $\gamma=0$ tend to have loops with a small Euclidean length and easily fall apart into disjoint components if one sequentially removes edges. With a large $\gamma$, generated networks are expected to be more robust against edge removal because they tend to have loops with a large Euclidean length, which generally offer multiple paths for many pairs of nodes.

\subsection{Examples of generated networks}

\begin{figure}[p]
  \centering
    \includegraphics[width=\textwidth]{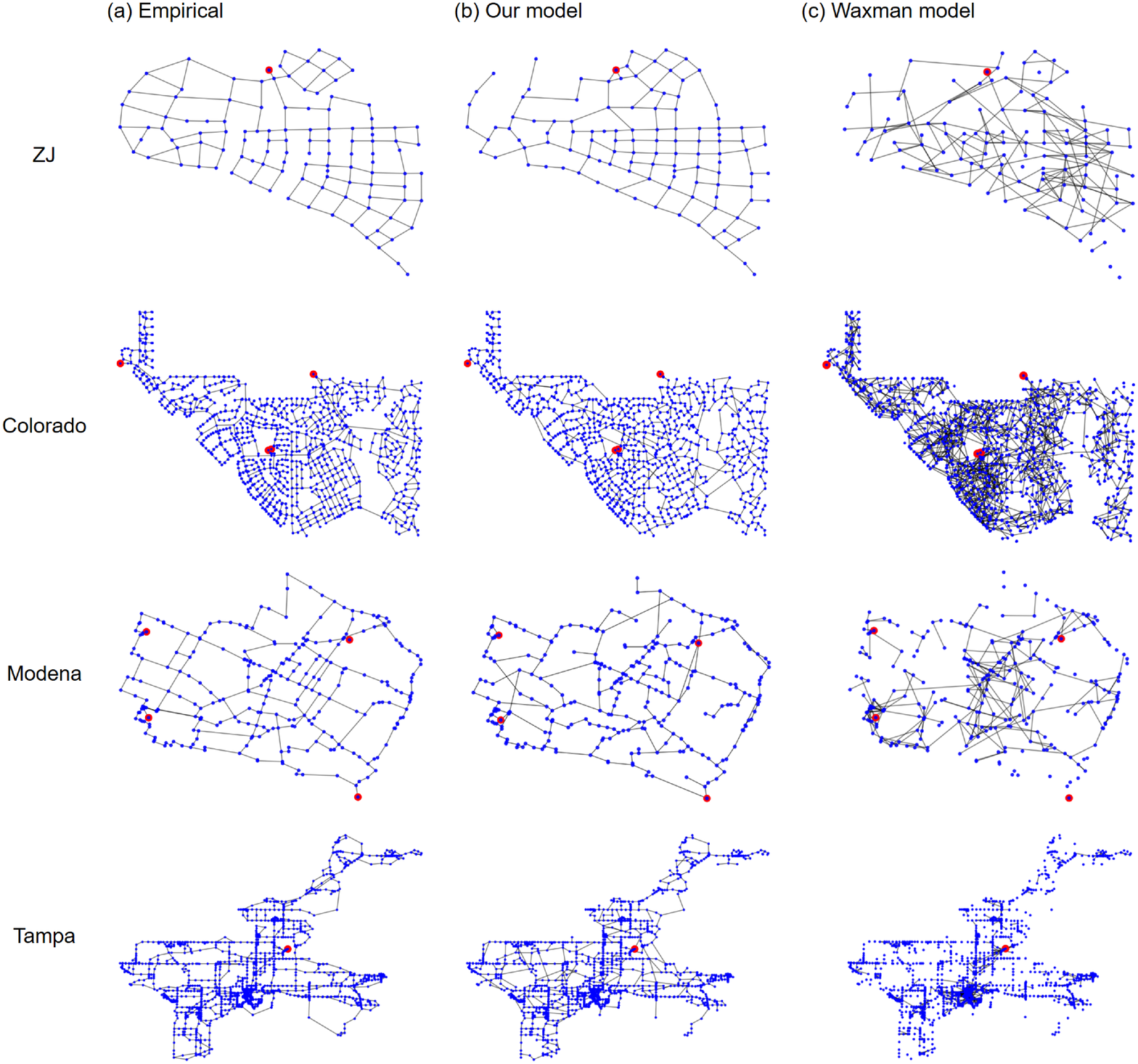}
    \caption{Visualization of the networks. The red and blue nodes represent the root nodes and the demand nodes, respectively. (a) Empirical WDNs. (b) Synthetic networks generated by our growth model. We set $\gamma=0.35$ for \textsc{ZJ} and \textsc{Tampa}, $\gamma=0.4$ for \textsc{Colorado}, and $\gamma=0.5$ for \textsc{Modena}. (c) Synthetic networks generated by the Waxman model. We set $\beta=0.619$ for \textsc{ZJ}, $\beta=0.193$ for \textsc{Colorado}, $\beta=0.596$ for \textsc{Modena}, and $\beta=0.256$ for \textsc{Tampa}.}
\label{fig:network_visualization}
\end{figure}

We visualize the networks that our growth model generates with $\gamma=0.35$ for \textsc{ZJ} and \textsc{Tampa}, $\gamma=0.4$ for \textsc{Colorado}, and $\gamma=0.5$ for \textsc{Modena} in Fig.~\ref{fig:network_visualization}(b). The synthetic networks are apparently similar to the empirical WDNs (Fig.~\ref{fig:network_visualization}(a)). For comparison, we also visualize the synthetic networks generated by the Waxman model, which is a spatial variant of random graphs (see Appendix \ref{appendix:Waxman} for details of the Waxman model), in Fig.~\ref{fig:network_visualization}(c). The Waxman model produces networks that are apparently dissimilar to the empirical WDNs.

We compare structural properties among the empirical WDNs, our model, and the Waxman model in Table \ref{tab:structural_properties_comparison}. The synthetic networks generated by our model are roughly similar to the empirical WDNs in terms of the indices measured. The synthetic networks generated by the Waxman model are not connected as one component, and therefore we do not compute the route factor and the robustness for them. Apart from the route factor and robustness, the Waxman model is roughly as similar to the empirical WDNs as our model is in terms of the average degree and meshedness coefficient. However, the Waxman model is substantially more dissimilar to the empirical WDNs than our model is in terms of the Jaccard index, maximum degree, and cost. 

\begin{figure}[p]
  \centering
    \includegraphics[width=\textwidth]{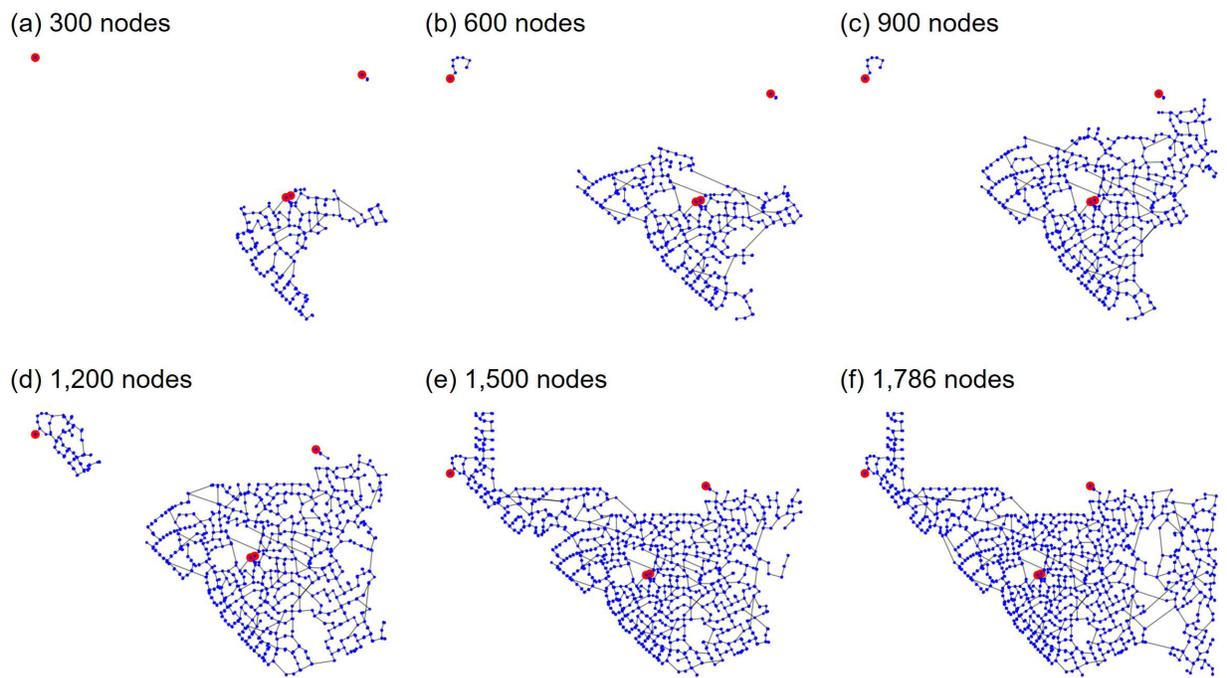}
    \caption{Visualization of the network growth in the case of \textsc{Colorado}. We simulated our network growth model with $\gamma=0.4$. For visual clarity, we omitted the isolated demand nodes.}
\label{fig:Visualization_Colorado_growth}
\end{figure}

We show a simulated time course of network growth with our model in the case of \textsc{Colorado} in Fig.~\ref{fig:Visualization_Colorado_growth}. We observe that the connected component associated with each root node expands its service area, and such connected components merge at later times.

\subsection{Trade-offs between the cost, route factor, and robustness}
\begin{figure}[p]
  \centering
    \includegraphics[width=\textwidth]{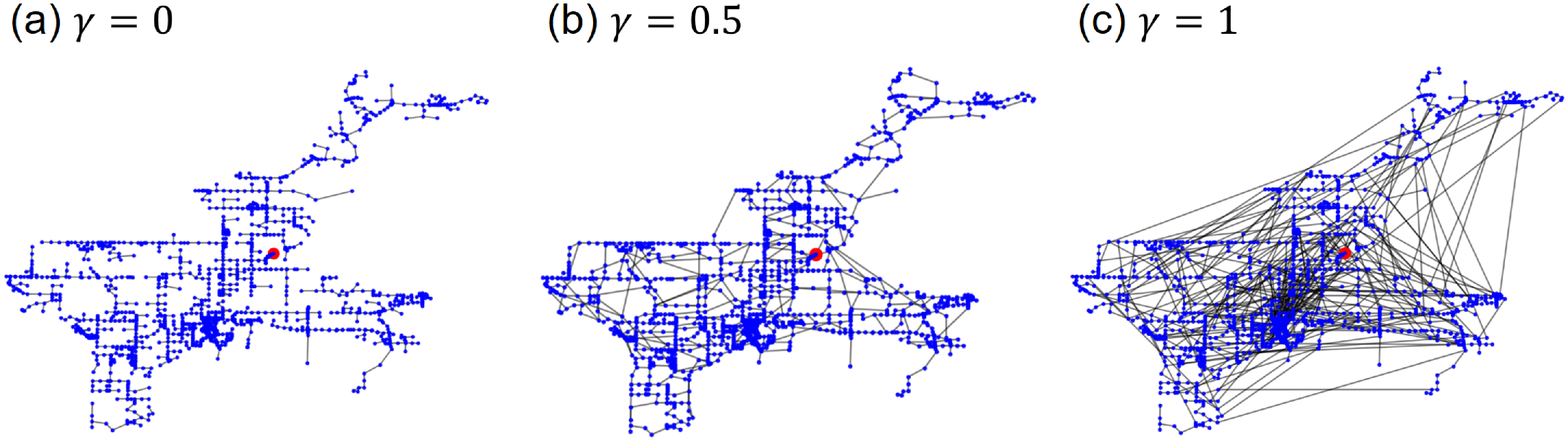}
    \caption{Visualization of synthetic networks generated by our model in the case of \textsc{Tampa}. (a) $\gamma=0$. (b) $\gamma=0.5$. (c) $\gamma=1$.}
\label{fig:different_gamma}
\end{figure}

Our model has parameter $\gamma$, which controls the trade-off between the wiring cost and the length of the added loops. We visualize sample synthetic networks generated by our model with $\gamma=0$, $\gamma=0.5$, and $\gamma=1$ in the case of \textsc{Tampa} in Fig.~\ref{fig:different_gamma}. With $\gamma=0$, we sequentially add the edges that yield the smallest Euclidean distance in each step. As shown in Fig.~\ref{fig:different_gamma}(a), the resulting network has loops with small Euclidean lengths. For a larger $\gamma$, there are more long edges that close loops. We computed the cost, route factor, and robustness for these three networks. The cost increases as $\gamma$ increases, i.e., it is equal to 367, 720, and 2{,}248 when $\gamma=0$, $\gamma=0.5$, and $\gamma=1$, respectively. The route factor decreases as $\gamma$ increases, i.e., it is equal to 2.20, 1.24, and 1.23 when $\gamma=0$, $\gamma=0.5$, and $\gamma=1$, respectively. The robustness increases as $\gamma$ increases, i.e., it is equal to 0.0184, 0.217, and 0.280 when $\gamma=0$, $\gamma=0.5$, and $\gamma=1$, respectively. 
This preliminary result indicates that there are trade-offs between the cost, route factor, and robustness as we vary $\gamma$. In this section, we study these trade-offs.

\begin{figure}[p]
  \centering
    \includegraphics[width=\textwidth]{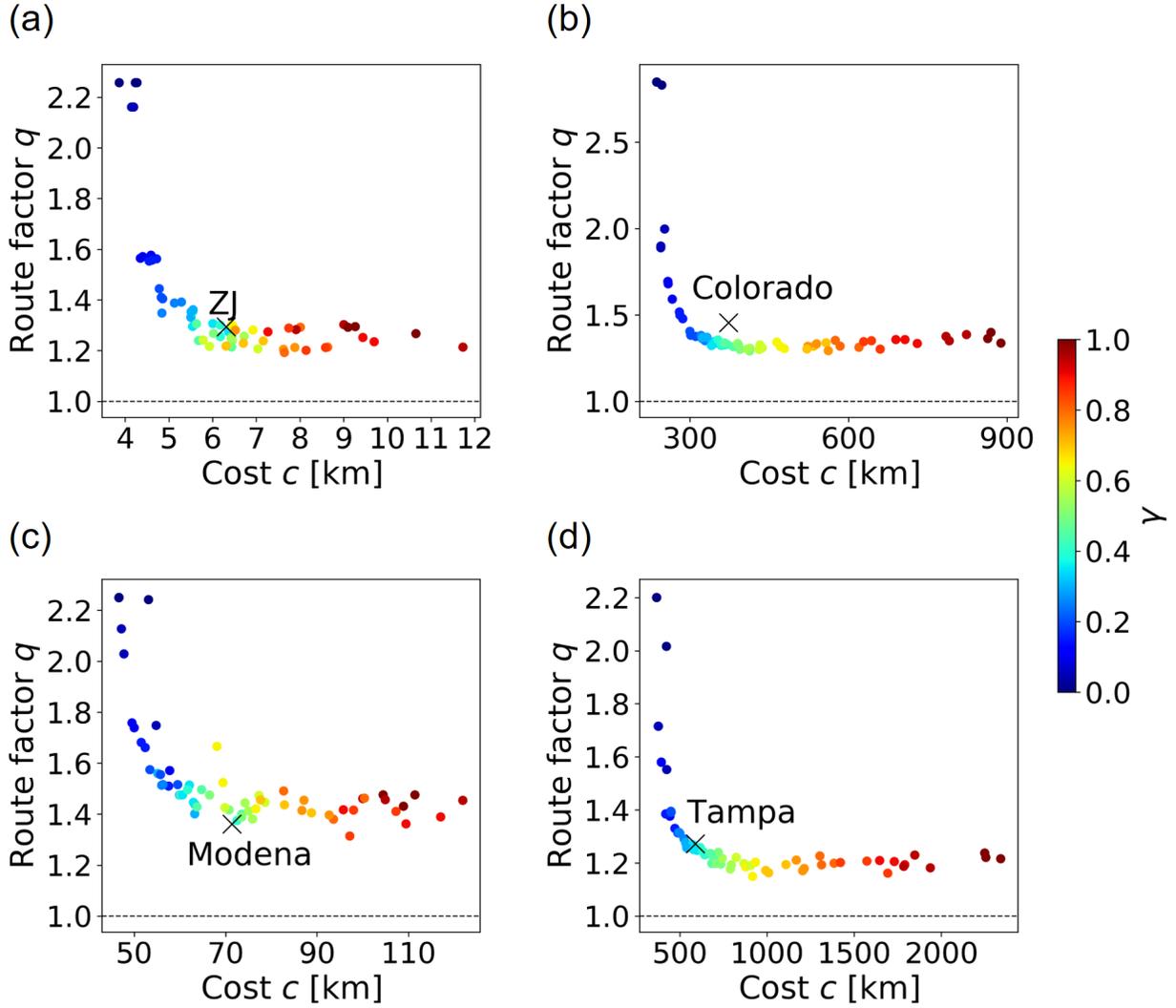}
\caption{Relationships between the cost and route factor. (a) \textsc{ZJ}. (b) \textsc{Colorado}. (c) \textsc{Modena}. (d) \textsc{Tampa}. The crosses represent the empirical WDNs. The circles represent synthetic networks. For the synthetic networks, we use $\gamma \in \{0, 0.05, \ldots, 1 \}$ and generate three networks for each value of $\gamma$. The dashed lines represent $q=1$.}
\label{fig:cost_routefactor}
\end{figure}

We start by comparing the cost and route factor for the empirical and synthetic networks. For the synthetic networks, we use $\gamma \in \{0, 0.05, \ldots, 1 \}$ and generate three networks for each value of $\gamma$. We show the results in Fig.~\ref{fig:cost_routefactor}. First, the cost tends to increase as $\gamma$ increases. This result is consistent with our observation with Fig.~\ref{fig:different_gamma}. Second, the route factor is the largest when $\gamma=0$, and it decreases sharply as $\gamma$ increases. The route factor is largely independent of $\gamma$ when $\gamma$ is approximately larger than 0.5. 
Third, our growth model generates networks similar to the empirical WDNs in terms of the cost and route factor when $0.35 \leq \gamma \leq 0.5$. The synthetic networks when $0.35 \leq \gamma \leq 0.5$ realize route factor values that are close to the minimum (i.e., less than $\approx 115\%$ of the smallest possible value when one varies $\gamma$) and a cost that is less than $\approx 170\%$ of the minimum for each empirical WDN.

\begin{figure}[p]
  \centering
    \includegraphics[width=\textwidth]{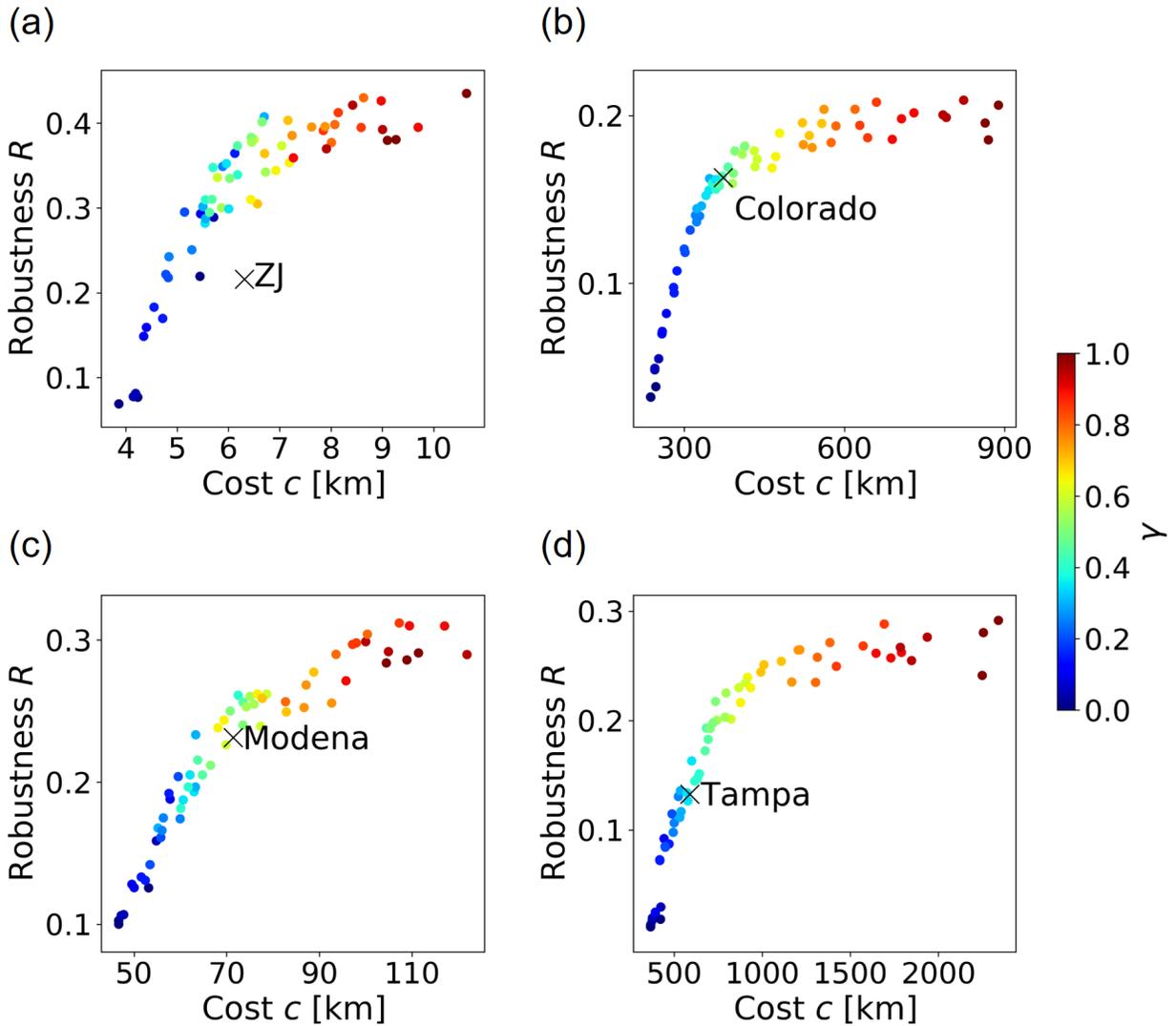}
    \caption{Relationships between the cost and robustness. (a) \textsc{ZJ}. (b) \textsc{Colorado}. (c) \textsc{Modena}. (d) \textsc{Tampa}. The crosses represent the empirical WDNs. The circles represent synthetic networks. For the synthetic networks, we use $\gamma \in \{0, 0.05, \ldots, 1 \}$ and generate three networks for each value of $\gamma$.}
\label{fig:cost_robustness}
\end{figure}

We show in Fig.~\ref{fig:cost_robustness} the trade-offs between the cost and robustness for synthetic networks across different values of $\gamma$. Again, for the synthetic networks, we use $\gamma \in \{0, 0.05, \ldots, 1 \}$ and generate three networks for each value of $\gamma$. The robustness increases as a function of $\gamma$, and it does so considerably more when $\gamma$ is small than when $\gamma$ is large. 
Similar to the case of the route factor, our growth model generates networks similar to the empirical WDNs in terms of the cost and robustness when $0.35 \leq \gamma \leq 0.5$ except for the case of \textsc{ZJ}. The synthetic networks with $0.35 \leq \gamma \leq 0.5$ realize robustness values that are more than $\approx 60\%$ of the maximum and a cost that is less than $\approx 170\%$ of the minimum for each of the empirical WDNs. 
The robustness of \textsc{ZJ} is smaller than those of the synthetic networks with a similar cost.
A possible reason for this result is that the \textsc{ZJ} network apparently has a strong community structure (see Fig.~\ref{fig:network_visualization}(a)), which our model does not intend to reproduce. 


\section{Pipe Diameter}
\label{sec:pipe_diameter}

\subsection{Background and the Physarum model}

We have neglected the diameter of pipes. 
We visualize the pipe diameters for the \textsc{Colorado}, \textsc{Modena}, and \textsc{Tampa} networks in Fig.~\ref{fig:diameters}(a). We exclude the \textsc{ZJ} network because the diameter of all the pipes in the \textsc{ZJ} network is 600 [cm]\cite{ZJ}. The \textsc{Modena} network has five different diameters
of pipes ranging from 100 to 400 [cm]. The \textsc{Colorado} network has eight different diameters of pipes ranging from 6 to 24 [inch]. The \textsc{Tampa} network has 34 different diameters of pipes ranging
from 8 to 54 [inch]. Figure~\ref{fig:diameters}(a) indicates that large-diameter pipes tend to be physically close to a root node. However, this is not always the case. For example, there are large-diameter pipes
that are located far from the root node in the upper part of the \textsc{Tampa} network. The diameter of pipes influences the installation cost and the performance of WDNs such as the flow velocity and the
head loss \cite{caballero2019water}. Abrupt changes in the pipe diameter may cause rapid variation in head loss, which may lead to physical damages and failures in pipes \cite{walski2001water}, and the gradual change of pipe diameters along flow paths has been shown to increase both energy and path redundancies in WDNs  \cite{abdel2021quantifying}. Therefore, the distribution of pipe diameters is an indispensable component for assessing the performance and design of WDNs.

Likewise, edge conductance per unit length (i.e., the reciprocal of the edge's resistance per unit length), which we call conductance for short in the following text, is an important determinant for various natural flow networks such as leaf veins of plants \cite{blonder2011venation}, vascular systems of animals \cite{gafiychuk2001principles}, and river networks \cite{rinaldo2014evolution}. These networks continuously adapt to the environment and are survivors of evolution. Therefore, they may attain optimal structure according to an evolutionary criterion. It is of interest to design transportation networks inspired by such natural networks. Models inspired by an amoeba-like organism, \textit{Physarum polycephalum}, have been used for designing optimal transportation networks \cite{liu2012physarum,tero2006physarum,tero2007mathematical, tero2010rules,zhang2014rapid}. 
These models operationalize positive feedback between the conductance of edges and the amount of flow passing through them.
This feedback mechanism is informed by the physiology of \textit{Physarum polycephalum} with which plasmodial tubes thicken if the protoplasmic flow through them increases \cite{nakagaki2000interaction}. 
In this section, we compare the pipe diameters for the empirical WDNs and the ones determined by a model \cite{tero2010rules}, which we refer to as the Physarum model. We use this specific model because the networks generated by this model have been suggested to be optimal in terms of the balance between the cost, efficiency, and robustness \cite{tero2010rules}.

Here we briefly describe the Physarum model proposed in Ref.~\cite{tero2010rules}. 
Under the assumption that the flow is laminar and follows the Hagen-Poiseuille equation, the flux through pipe $(i,j)$, denoted by $Q_{ij}$, is given by
\begin{equation}
    Q_{ij} = \frac{\pi R_{ij}^{4}(p_i-p_j)}{128\eta \ell_{ij}}=\frac{D_{ij}(p_i-p_j)}{\ell_{ij}}\,,
    \label{eq:Physarum_Q}
\end{equation}
where the flow is measured in the direction from the $i$th to $j$th nodes; $R_{ij}$ is the diameter of pipe $(i, j)$; $\eta$ is the viscosity of the fluid; $D_{ij}=\pi R_{ij}^{4}/128 \eta$ is the conductance of pipe $(i, j)$; $\ell_{ij}$ is the physical length of pipe $(i, j)$; and $p_i$ is the pressure at node $i$. We consider the conservation law of flux given by
\begin{equation}
    \sum_{j=1}^M Q_{ij} = \sum_{j=1}^M\frac{D_{ij}(p_i-p_j)}{\ell_{ij}}= \left\{ \begin{array}{ll}
    I_0/|V_R| & (i\in V_{R})\,, \\
    -I_0/|V_D| & (i\in V_{D})\,,
  \end{array} \right.
    \label{eq:Physarum_conservation}
\end{equation}
where $I_0$ (with $I_0 > 0$) is the total flux outgoing from the root nodes. We remind that $V_R$ and $V_D$ are the set of the root nodes and that of the demand nodes, respectively, such that $V = V_R \cup V_D$. Equation \eqref{eq:Physarum_conservation} assumes that all the root nodes have the same flux value and that all the demand nodes have the same flux value.

The conductance evolves in time according to
\begin{equation}
    \frac{{\rm d}D_{ij}}{{\rm d}t} = f(|Q_{ij}|)-D_{ij}\,.
    \label{eq:Physarum_Q_delta}
\end{equation}
We assume that the viscosity of the fluid $\eta$ is constant, which implies that the diameter $R_{ij}$ evolves in proportion to $\left(D_{ij}\right)^{1/4}$ as $D_{ij}$ evolves. The first term on the right-hand side of Eq.~\eqref{eq:Physarum_Q_delta} describes the increase in $D_{ij}$ in response to the flux, which is a positive feedback effect. In the absence of flow, $D_{ij}$ exponentially decays over time because of the second term.
We use
\begin{equation}
    f(|Q_{ij}|) = \frac{|Q_{ij}|^{\mu}}{1+|Q_{ij}|^{\mu}}\,,
    \label{eq:Physarum_fQ}
\end{equation}
where $\mu(>0)$ is a parameter that specifies the nonlinearity of the positive feedback. 

\subsection{Numerical results}
We apply the Physarum model to the three empirical WDNs (i.e., \textsc{Colorado}, \textsc{Modena}, and \textsc{Tampa}) and the corresponding synthetic networks generated by our growth model. The synthetic networks are those shown in Fig.~\ref{fig:network_visualization}(b).

\begin{figure}[p]
  \centering
    \includegraphics[width=\textwidth]{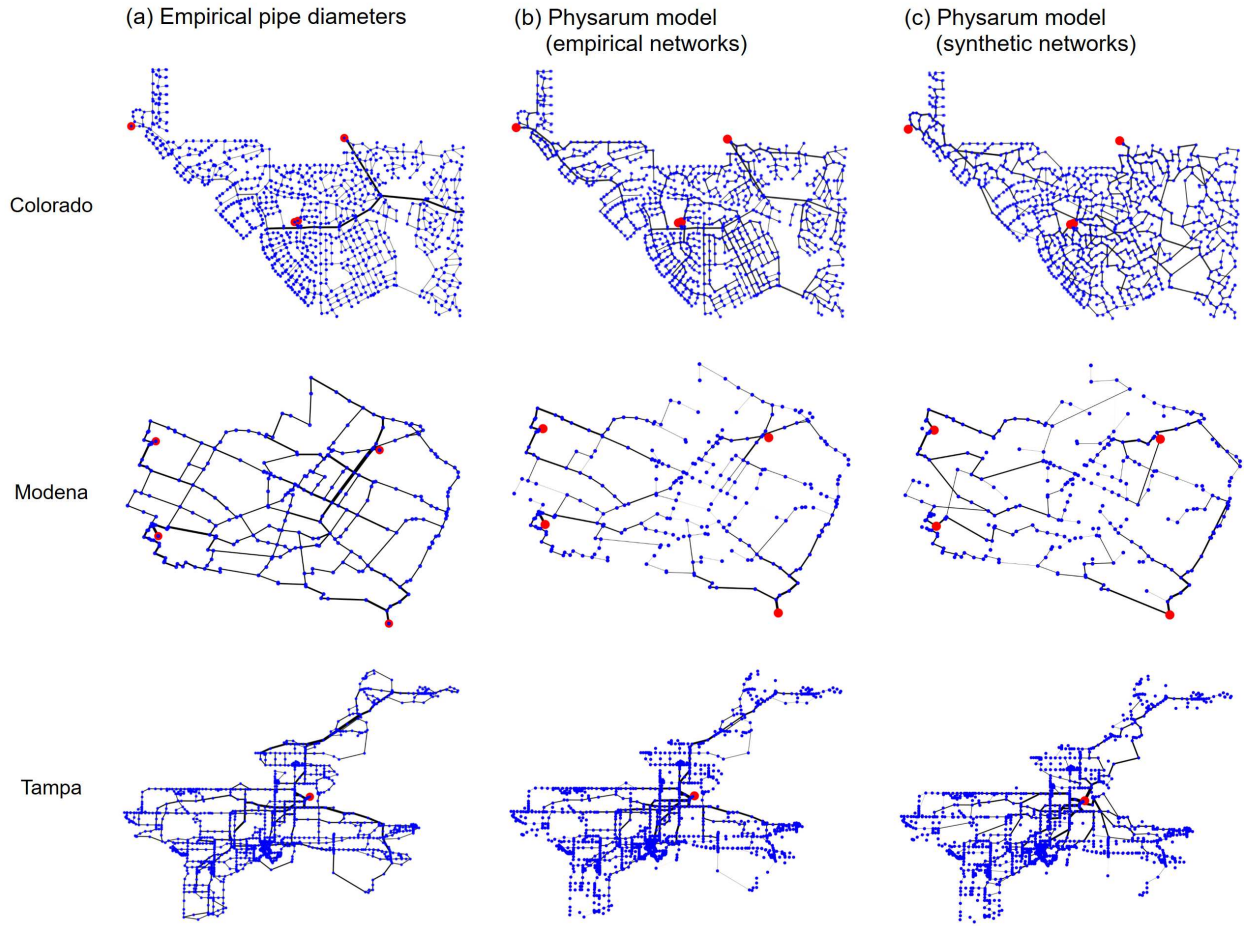}
    \caption{Visualization of the networks with pipe diameters. (a) Empirical networks with the empirical pipe diameters. (b) Empirical networks with the pipe diameters that are determined by the Physarum model. (c) Synthetic networks generated by our growth model in which the pipe diameters are determined by the Physarum model. The synthetic networks are the ones we showed in Fig.~\ref{fig:network_visualization}(b). In the Physarum model used in (b) and (c), we set $\mu=1$ and $I_0=10$.}
\label{fig:diameters}
\end{figure}

We visualize the networks with the pipe diameters, $R_{ij}$, being obtained by the Physarum model with $\mu=1$ and $I_0=10$ for the empirical and synthetic networks in Figs.~\ref{fig:diameters}(b) and \ref{fig:diameters}(c), respectively. In these figures, there are large-diameter pipes physically close to the root nodes, which is similar to the empirical WDNs shown in Fig.~\ref{fig:diameters}(a). In addition, the Physarum model generates large-diameter pipes in the upper part of the \textsc{Tampa} network (see Figs.~\ref{fig:diameters}(b) and \ref{fig:diameters}(c)), which is also consistent with the empirical \textsc{Tampa} network (see Fig.~\ref{fig:diameters}(a)). However, there are also dissimilarities especially in the case of the synthetic networks shown in Fig.~\ref{fig:diameters}(c). For example, the distribution of pipe diameters in \textsc{Colorado} in Fig.~\ref{fig:diameters}(c) is not close to that of the WDN shown in Fig.~\ref{fig:diameters}(a). 

\begin{figure}[p]
  \centering
    \includegraphics[width=\textwidth]{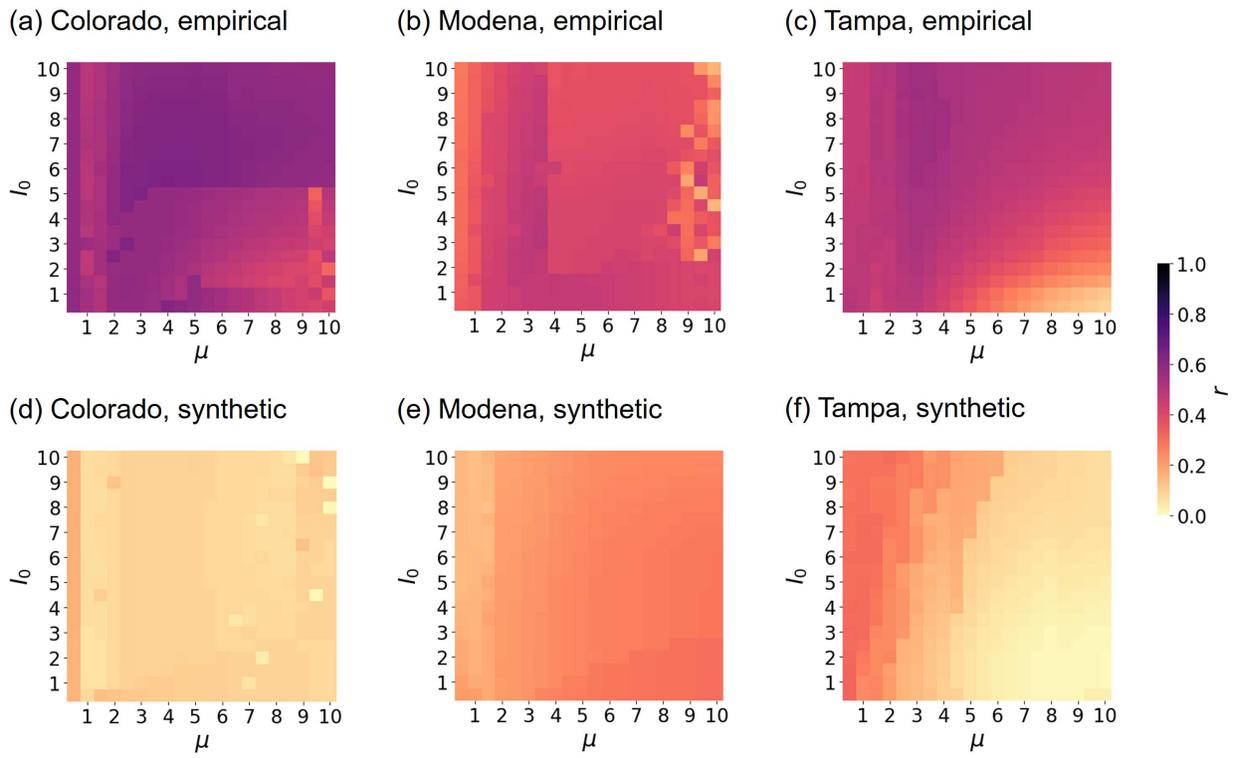}
    \caption{Pearson correlation coefficient, $r$, between the empirical and simulated pipe diameters.}
\label{fig:heatmap}
\end{figure}

To quantify the similarity between the empirical and synthetic networks in terms of the distribution of the pipe diameters, we measure the Pearson correlation coefficient between the pipe diameter obtained by the Physarum model and the empirical counterpart, where we regard each pipe as a sample. In the case of synthetic networks, we calculate the Pearson correlation coefficient only based on the edges co-present in the empirical and synthetic networks. 
We show the Pearson correlation coefficient, denoted by $r$, for different values of the two parameters of the Physarum model, $\mu$ and $I_0$, in Fig.~\ref{fig:heatmap}. According
to a standard, we interpret the Pearson correlation coefficient to be strong, moderate, or weak when $|r| > 0.7$, $|r| > 0.4$, or $|r| > 0.1$, respectively \cite{dancey2007statistics}. Figure~\ref{fig:heatmap} suggests that, given the structure of the empirical network, the correlation between the empirical data and the synthetic data obtained from the Physarum model is moderately positive albeit not strongly positive in a broad parameter region for all the three WDNs. In contrast, the pipe diameter for the synthetic networks is only weakly correlated with the empirical data.

\section{Conclusions}
\label{sec:conclusions}
We proposed a growth model for WDNs, which does not require detailed inputs apart form the physical location of each node. We have found that our model produces networks whose structure is similar to that of the empirical WDNs when $0.35 \leq \gamma \leq 0.5$. These networks realize route factor values that are less than $\approx 115\%$ of the smallest possible value when one varies $\gamma$, robustness values that are more than $\approx 60\%$ of the largest possible value, and cost values that are less than $\approx 170\%$ of the smallest possible value for each of the four empirical WDNs that we have used. 
When $\gamma \leq 0.35$, the route factor tends to decrease drastically as one increases $\gamma$. On the other hand, when $\gamma \geq 0.5$, the gain in the robustness when one increases $\gamma$ is small. Therefore, the empirical WDNs may realize a reasonable balance between the cost, efficiency, and robustness. These results imply that our model may inform growth mechanisms of real-world WDNs and design of WDNs. Because the range of the $\gamma$ values for which the synthetic network is most similar to the empirical WDN is close for the four empirical WDNs (i.e., $0.35 \leq \gamma \leq 0.5$), the proposed growth model with this range of $\gamma$ may be able to explain growth dynamics of various other real-world WDNs. The model may also be applicable to other spatial flow networks such as gas pipeline networks. Nonetheless, our model and its evaluation only depended on the network structure. Future studies along this line should examine  the hydraulic properties of WDNs for our model to be informative in practical contexts.

Some studies showed that increasing degree-degree correlations is effective at improving the network robustness  \cite{schneider2011mitigation,tanizawa2012robustness,wu2011onion}. There is also a strong relation between robustness and loops  \cite{braunstein2016network,chujyo2021loop,dziedzic2014water,hayashi2018onion,katifori2010damage}. In the present study, we have shown that the robustness is highly dependent on the value of $\gamma$ for $\gamma \leq 0.5$, which suggests that the distribution of loops may be a key determinant of the robustness. In Ref. \cite{katifori2012quantifying}, the authors proposed a method for analyzing the distribution of loops for planar graphs and found that leaf venation networks have hierarchical structure with which large loops contain small loops. Such a hierarchical structure of loops may allow leaves to maintain the supply of water and nutrients even when flow through some edges in the network is lost due to damages \cite{ronellenfitsch2019phenotypes,ronellenfitsch2015topological,sack2008leaf}. In addition, loop nestedness increases flow path redundancies in WDNs \cite{abdel2021quantifying}. Our results also show that such hierarchical structure of loops may exist for some values of $\gamma$ (see Fig.~\ref{fig:different_gamma}). However, analyzing the distribution of loops for non-planar graphs is still an open question \cite{mileyko2012hierarchical,modes2016extracting}. Further studies are desirable for investigating the correlation between the network-based robustness and hydraulic performance of WDNs, and clarifying relationships between the robustness and the distribution of loops in non-planar graphs including many of WDNs.

We also studied the design of pipe diameters based on a biological positive feedback mechanism. We found moderate positive correlations between the empirical and modeled pipe diameter across the pipes for all the three WDNs. These results suggest that the distribution of the empirical pipe diameters may be closer to optimal than to a uniformly random distribution under the assumption that the Physarum model approximately optimizes the conductance of edges in terms of some efficiency or robustness criteria. In fact, a similar Physarum model realizes distributions of pipe diameters that realize a small overall wiring cost in WDNs \cite{skulovichadaptation}. At the same time, the Physarum model or similar models lack hydraulic considerations, and it is underexplored whether optimality in the sense of evolution of amoeba-like organisms has anything to do with hydraulic performances. For example, the Physarum model does not take into account the elevation change over the pipe's length. When such elevation changes are nonnegligible, it may be necessary to adjust the model in this respect for practical applications. In addition, we assumed for simplicity that all the root nodes have the same flux value and that all the demand nodes have the same flux value  (see Eq.~\eqref{eq:Physarum_conservation}). Obviously, it is of pragmatic interest to adjust the model to take into account heterogeneity in these quantities. Further studies should also compare the hydraulic performance between the empirical WDNs and the networks whose pipe diameters are obtained by these biologically inspired models. Such studies will provide the insight on the fundamental contribution of network structure (i.e., edge configuration and pipe diameter arrangement) to performance of WDNs. Moreover, additional nature-inspired mechanisms such as the growth of underlying tissues in leaf venation networks \cite{ronellenfitsch2016global} and other optimization techniques such as simulated annealing \cite{katifori2010damage} and combination between biological principles and engineering control \cite{jiang2019optimizing} may also be useful for realizing desirable distributions of pipe diameters.

\appendix
\begin{appendices}
\section{Waxman model}
\label{appendix:Waxman}
The Waxman model is a spatial variant of the Erd\H{o}s–R\'{e}nyi model \cite{waxman1988routing}. In the original Waxman model, the nodes are uniformly distributed in the two-dimensional space. In contrast, we use the two-dimensional coordinate of each node informed by the empirical data. One adds each edge between each pair of nodes, $i$ and $j$, with probability
\begin{equation}
        P_{ij} = \beta e^{-d_{ij}/d_0}\,,
        \label{eq:waxman}
    \end{equation}
where $d_0$ is the typical length of edge, and $\beta$ controls the density of edges. One determines whether or not to lay an edge between each pair of the $i$th and $j$th nodes independently for different node pairs. We set $d_0$ to the average Euclidean length of edges in the given empirical WDN. For each WDN, we determine the $\beta$ value by $\sum_{i=1}^N \sum_{j=1}^{i-1}P_{ij}=M$, which guarantees that the expected number of edges generated by the Waxman model is equal to the number of edges in the empirical WDN. The obtained $\beta$ values are 0.619 for \textsc{ZJ}, 0.193 for \textsc{Colorado}, 0.596 for \textsc{Modena}, and 0.256 for \textsc{Tampa}. 
\end{appendices}

\enlargethispage{20pt}

\section*{Data Accessibility}
The empirical data sets for the \textsc{ZJ}, \textsc{Colorado}, and \textsc{Modena} networks are available at \cite{Colorado_Springs, Modena,ZJ}.

\section*{Authors' Contributions}
N.AM., Q.Z., and N.M. conceived the research. K.S. and N.M. developed the model. K.S. performed the numerical analysis. K.S., N.AM., Q.Z., and N.M. discussed the results and wrote the paper.

\section*{Competing Interests}
We declare we have no competing interests.

\section*{Funding}
K.S. thanks the financial support by the Japan Society for the Promotion of Science (under Overseas Research Fellowships and Grant No.~19K23531). N.AM. and Q.Z. thank the financial support by the National Science Foundation (under Grant No.~1638301). Any opinions, findings, and conclusions or recommendations expressed in this material are those of the authors and do not necessarily reflect the views of the National Science Foundation. N.M. thanks the financial support by AFOSR European Office (under Grant No.~FA9550-19-1-7024), the Nakatani Foundation, and the Sumitomo Foundation. 

\section*{Acknowledgements}
The authors thank Brian Pickard and Seung Park from the city of Tampa
Water Department for providing the water network data.









\bibliographystyle{abbrv}
\bibliography{References}

\clearpage

\clearpage

\end{document}